\documentclass[conference]{IEEEtran}
\IEEEoverridecommandlockouts
\usepackage{cite}
\usepackage{amsmath,amssymb,amsfonts}
\usepackage{graphicx}
\usepackage{textcomp}
\usepackage[table]{xcolor}
\usepackage{pgfplots,pgfplotstable}
\usepackage{subcaption}
\usepackage{booktabs}
\usepackage{pgf-umlsd}
\pgfplotsset{compat=1.18}
\usepgfplotslibrary{statistics}

\pgfplotsset{
    box plot/.style={
        /pgfplots/.cd,
        black,
        only marks,
        mark=-,
        mark size=\pgfkeysvalueof{/pgfplots/box plot width},
        /pgfplots/error bars/y dir=plus,
        /pgfplots/error bars/y explicit,
        /pgfplots/table/x index=\pgfkeysvalueof{/pgfplots/box plot x index},
    },
    box plot box/.style={
        /pgfplots/error bars/draw error bar/.code 2 args={%
            \draw  ##1 -- ++(\pgfkeysvalueof{/pgfplots/box plot width},0pt) |- ##2 -- ++(-\pgfkeysvalueof{/pgfplots/box plot width},0pt) |- ##1 -- cycle;
        },
        /pgfplots/table/.cd,
        y index=\pgfkeysvalueof{/pgfplots/box plot box top index},
        y error expr={
            \thisrowno{\pgfkeysvalueof{/pgfplots/box plot box bottom index}}
            - \thisrowno{\pgfkeysvalueof{/pgfplots/box plot box top index}}
        },
        /pgfplots/box plot
    },
    box plot top whisker/.style={
        /pgfplots/error bars/draw error bar/.code 2 args={%
            \pgfkeysgetvalue{/pgfplots/error bars/error mark}%
            {\pgfplotserrorbarsmark}%
            \pgfkeysgetvalue{/pgfplots/error bars/error mark options}%
            {\pgfplotserrorbarsmarkopts}%
            \path ##1 -- ##2;
        },
        /pgfplots/table/.cd,
        y index=\pgfkeysvalueof{/pgfplots/box plot whisker top index},
        y error expr={
            \thisrowno{\pgfkeysvalueof{/pgfplots/box plot box top index}}
            - \thisrowno{\pgfkeysvalueof{/pgfplots/box plot whisker top index}}
        },
        /pgfplots/box plot
    },
    box plot bottom whisker/.style={
        /pgfplots/error bars/draw error bar/.code 2 args={%
            \pgfkeysgetvalue{/pgfplots/error bars/error mark}%
            {\pgfplotserrorbarsmark}%
            \pgfkeysgetvalue{/pgfplots/error bars/error mark options}%
            {\pgfplotserrorbarsmarkopts}%
            \path ##1 -- ##2;
        },
        /pgfplots/table/.cd,
        y index=\pgfkeysvalueof{/pgfplots/box plot whisker bottom index},
        y error expr={
            \thisrowno{\pgfkeysvalueof{/pgfplots/box plot box bottom index}}
            - \thisrowno{\pgfkeysvalueof{/pgfplots/box plot whisker bottom index}}
        },
        /pgfplots/box plot
    },
    box plot median/.style={
        /pgfplots/box plot,
        /pgfplots/table/y index=\pgfkeysvalueof{/pgfplots/box plot median index}
    },
    box plot width/.initial=1em,
    box plot x index/.initial=0,
    box plot median index/.initial=1,
    box plot box top index/.initial=2,
    box plot box bottom index/.initial=3,
    box plot whisker top index/.initial=4,
    box plot whisker bottom index/.initial=5,
}

\newcommand{\boxplot}[2][]{
    \addplot [box plot median,#1] table {#2};
    \addplot [forget plot, box plot box,#1] table {#2};
    \addplot [forget plot, box plot top whisker,#1] table {#2};
    \addplot [forget plot, box plot bottom whisker,#1] table {#2};
}

\def\BibTeX{{\rm B\kern-.05em{\sc i\kern-.025em b}\kern-.08em
    T\kern-.1667em\lower.7ex\hbox{E}\kern-.125emX}}
\usepackage{comment}

\usepackage{fancyhdr} % Used for customizing the headers and footers, e.g., for adding copyright info to the author copy.
\usepackage{polaristools}
\usepackage[capitalise]{cleveref}
\usepackage[T1]{fontenc}    % Determines font encoding of the output. Font packages have to be included before this line.

\makeatletter
\newcommand{\linebreakand}{%
  \end{@IEEEauthorhalign}
  \hfill\mbox{}\par
  \mbox{}\hfill\begin{@IEEEauthorhalign}
}
\makeatother

\ifthenelse{\boolean{anonymousMode}}{
    \renewcommand{\thanks}[1]{}
}{
}

\fancypagestyle{fancycopyright}{%
    \fancyhf{}%
    \fancyhead[C]{Author copy of paper published at \emph{2025 IEEE International Conference on Edge Computing and Communications (EDGE)}\\
    \copyright2025 IEEE% Official publication: \url{https://doi.org/10.1109/CLOUD53861.2021.00085}
    }%
    \fancyfoot[C]{\thepage}%
}

\begin{document}

% IEEE template: https://www.ieee.org/conferences/publishing/templates.html

\title{Stardust: A Scalable and Extensible Simulator\\ for the 3D Continuum
    % \thanks{This project has received funding from the European Union’s Horizon 2020 research and innovation programme under grant agreement No 871403.}
}

\ifthenelse{\boolean{anonymousMode}}{
    \author{\IEEEauthorblockN{Anonymous Authors}
    \IEEEauthorblockA{}
    }
}{
    \author{
        \IEEEauthorblockN{Thomas Pusztai}
        \IEEEauthorblockA{\textit{Distributed Systems Group, TU Wien} \\
            t.pusztai@dsg.tuwien.ac.at
        }
        \and
        \IEEEauthorblockN{Jan Hisberger}
        \IEEEauthorblockA{\textit{Distributed Systems Group, TU Wien} \\
            e12126323@student.tuwien.ac.at
        }
        \and
        \IEEEauthorblockN{Cynthia Marcelino}
        \IEEEauthorblockA{\textit{Distributed Systems Group, TU Wien} \\
            c.marcelino@dsg.tuwien.ac.at
        }
        \linebreakand
        \IEEEauthorblockN{Stefan Nastic}
        \IEEEauthorblockA{\textit{Distributed Systems Group, TU Wien} \\
            snastic@dsg.tuwien.ac.at}
    }
}

% Acronym and other definitions are placed in this file.

\newacronym{IoT}{IoT}{Internet of Things}
\newacronym{LEO}{LEO}{Low Earth Orbit}
\newacronym{EO}{EO}{Earth observation}
\newacronym{ISL}{ISL}{inter-satellite laser link}
\newacronym{OEC}{OEC}{Orbital Edge Computing}

\newcommand{\LEO}[0]{\gls{LEO}}
\newcommand{\EO}[0]{\gls{EO}}
\newcommand{\ISL}[0]{\gls{ISL}}
\newcommand{\ISLs}[0]{\glspl{ISL}}

% Frequently used inline code keywords.
\newcommand{\IRouter}[0]{\inlinecode{IRouter}}
\newcommand{\ILink}[0]{\inlinecode{ILink}}
\newcommand{\Computing}[0]{\inlinecode{Computing}}
\newcommand{\Node}[0]{\inlinecode{Node}}

\ifthenelse{\boolean{anonymousMode}}{
    \newcommand{\PolarisShortName}[0]{OurCloud}
    \newcommand{\PolarisLongProjectName}[0]{``Name omitted for double-blind review''}
    
    \newcommand{\PolarisProjectUrl}[0]{URL omitted for double-blind review}
    \newcommand{\StardustRepoUrl}[0]{https://github.com/<anonymized> and \url{https://doi.org/10.5281/zenodo.15002274}}    
}{
    \newcommand{\PolarisShortName}[0]{Polaris}
    \newcommand{\PolarisLongProjectName}[0]{Polaris SLO Cloud}
    
    \newcommand{\PolarisProjectUrl}[0]{\url{https://polaris-slo-cloud.github.io}}
    \newcommand{\StardustRepoUrl}[0]{\url{https://github.com/polaris-slo-cloud/stardust} and \url{https://doi.org/10.5281/zenodo.15484629}}    
}

\newcommand{\Polaris}[0]{\PolarisShortName{}}

% Names and numbers that may be changed later
\newcommand{\MaxEvalNodes}[0]{20.6k}

\maketitle

%%%%% Uncomment for page numbers (not part of official IEEE template)
% \thispagestyle{plain}
% \pagestyle{plain}

%%%%% Uncomment for page numbers and copyright info (for author copy)
\thispagestyle{fancycopyright}
\pagestyle{fancy}

\begin{abstract}
Low Earth Orbit~(LEO) satellite constellations are quickly being recognized as an upcoming extension of the Edge-Cloud Continuum into a 3D Continuum.
Low-latency connectivity around the Earth and increasing computational power with every new satellite generation lead to a vision of workflows being seamlessly executed across Edge, Cloud, and space nodes.
High launch costs for new satellites and the need to experiment with large constellations mandate the use of simulators for validating new orchestration algorithms.
Unfortunately, existing simulators only allow for relatively small constellations to be simulated without scaling to a large number of host machines.
In this paper, we present Stardust, a scalable and extensible simulator for the 3D Continuum.
Stardust supports
\begin{enumerate*}[i)]
    \item simulating mega constellations with 3x the size of the currently largest LEO mega constellation on a single machine,
    \item experimentation with custom network routing protocols through its dynamic routing mechanism, and
    \item rapid testing of orchestration algorithms or software by integrating them into the simulation as SimPlugins.
\end{enumerate*}
We evaluate Stardust in multiple simulations to show that it is more scalable than the state-of-the-art and that it can simulate a mega constellation with up to \MaxEvalNodes{}~\hlRevB{satellites} on a single machine.
\end{abstract}

\begin{IEEEkeywords}
3D continuum, edge-cloud continuum, orbital edge computing, LEO satellites, simulator
\end{IEEEkeywords}

% (Thomas optional) Reduce the gap between floats and text
% \setlength{\textfloatsep}{5pt}
% \setlength{\dbltextfloatsep}{5pt}

%%%%%%%%%%%%%%%%%%%%%%%%%%%%%%%%%%%%%%%%%%%%%%
% Include the sections of the paper here.
%%%%%%%%%%%%%%%%%%%%%%%%%%%%%%%%%%%%%%%%%%%%%%
\section{Introduction}
\label{Stardust:sec:Intro}

\LEO{} satellites are rapidly increasing in number in recent years.
As of 2024, there are more than 8,000 \LEO{} satellites in orbit~\cite{LEO_SAT_stats}, with about 7,000~of these belonging to the Starlink mega constellation~\cite{FCC_StarlinkCompetition2024}, which plans to grow to more than 12,000~satellites by 2028~\cite{StarlinkLowerOrbitAuth2019}.
Amazon intends to have its Kuiper mega constellation with more than 3,236~satellites~\cite{FCC_KuiperAuthorization2020} complete by 2029 and the FCC has called for more competition in this sector~\cite{FCC_StarlinkCompetition2024}.

\LEO{} mega constellations provide low latency communication between \LEO{} and terrestrial nodes and among terrestrial nodes.
For example, Starlink's median client-LEO-Cloud round-trip latency has recently been measured to be 40-50~ms~\cite{StarlinkPerf2024}.
The low altitude of \LEO{} satellites compared to geostationary satellites allows for low latency with terrestrial nodes that are directly in range.
\Glspl{ISL} enable the creation of large orbital networks~\cite{NetwTopology27000Kmh2019}, with \ISL{} speeds demonstrated up to 100~Gbps~\cite{Geo2LeoSat100Gbps2024}.

Since \LEO{} satellites get more computational capabilities with every new generation, several uses beyond bent pipe communication are being investigated.
Processing of \EO{} data on a single satellite has been shown by ESA~\cite{WildRide2023} and plans for processing data in clusters of satellites have been proposed under various names, such as \gls{OEC}~\cite{OECNano2020,InOrbitComputing2020}, satellite computing~\cite{Earth2Space2023,L2D2_2021}, or Edge-Cloud-Space 3D Continuum~\cite{HyperDrive2024,Cosmos2025}.
Possible use cases include federated learning in space~\cite{FLOEC2024,SatelliteBasedFL2022,FL_LEO_Clustering2023,optimizingFLscheduling2023}, \EO{} data compression for efficient downlinking~\cite{Fool2025}, smart agriculture~\cite{SatIoTSmartAgri20223}, and disaster response~\cite{HyperDrive2024}.
The actual use of satellites varies between the ideas, from preprocessing \EO{} data to full-fledged compute nodes that enable seamless execution of workloads across the 3D Continuum.
All proposals have in common that they require ways to evaluate their designs.

Since launching new satellites is expensive and the computing capabilities anticipated for the near future are not available in space yet, evaluation of \LEO{} computing systems must be performed using simulators or emulators.
Various simulators already exist, e.g., Hypatia~\cite{Hypatia2020}, Celestial~\cite{Celestial2022}, and StarryNet~\cite{StarryNet2023}.
While all of them compute satellite trajectories and node-to-node latencies, most simulators have a number of shortcomings.
Many solutions are emulators, e.g., Celestial~\cite{Celestial2022} and StarryNet~\cite{StarryNet2023}, which execute microVMs or containers for each node of the 3D~Continuum.
This has the advantage of enabling tests of real software systems, but executing one microVM or container for every node, regardless of whether it is used or not, limits the maximum infrastructure size that can be evaluated due to the resource usage.
Celestial allows suspending all VMs that are not in a certain area, but it still creates a microVM for each node.
Conversely, simulators, such as Hypatia~\cite{Hypatia2020}, do not execute nodes and are used for network simulation only, hence software testing or evaluation of placement algorithms that require node resource information are not possible.
Many simulators and emulators focus on the network latencies, precompute them before the experiment, and disregard the positions of satellites during the experiment.
Hence, it is often impossible to determine where a particular node is located during the experiment, although this information may be needed, e.g., for location-aware scheduling or for picking a satellite that will be in range of a certain ground station when it completes the next workload.
Many solutions account only for \LEO{} and ground station or Cloud nodes, but not for terrestrial Edge nodes, such as drones.
Typically, each simulator or emulator is designed for a single purpose only, e.g., testing software under resource and network constraints or evaluating network routing algorithms.
Existing solutions often lack extensibility, such as allowing custom logic to execute after every simulation step, which could, e.g., be used to add a new deployment to the experiment.

In this paper, we present Stardust, a scalable and extensible open-source\footnote{\StardustRepoUrl{}} simulator for the 3D Continuum.
Our main contributions are:
\begin{enumerate}
    \item \emph{Stardust, a scalable and extensible next generation simulator} for the 3D Continuum with support for simulating \LEO{}-, Cloud-, and Edge nodes in a scalable manner.
    Stardust enables experiments for evaluating networking and orchestration algorithms for the 3D Continuum.
    It supports simulating mega constellations three times the size of the currently largest constellation, with almost 7k~satellites on a single machine.

    \item A \emph{dynamic routing mechanism} that enables experimentation with different routing mechanisms by making the \ISL{} routing protocol and the network path computation changeable.
    This allows, e.g., changing the default +Grid \ISL{} routing to a different protocol or to introduce caching or hypergraph algorithms as a replacement for Dijkstra's algorithm to calculate node-to-node network paths.

    \item \emph{SimPlugin, a plugin mechanism} that serves as the integration point for custom logic that Stardust should execute at every step of the simulation.
    A SimPlugin has access to the complete infrastructure state and, thus, allows integrating, e.g., orchestration algorithms/software that should be evaluated using Stardust.
\end{enumerate}

The rest of this paper is structured as follows: \cref{Stardust:sec:Motivation} presents a motivating use case and requirements for a next-generation simulator for the 3D Continuum.
\cref{Stardust:sec:RelWork} explores other 3D Continuum simulators and emulators and \cref{Stardust:sec:Design} presents the design of the Stardust simulator. %, and \cref{Stardust:sec:Impl} its implementation.
In \cref{Stardust:sec:Evaluation} we evaluate Stardust in multiple simulations and in \cref{Stardust:sec:Conclusion} we conclude the paper and present future work.

\section{Motivating Use Case \& Simulator Requirements}
\label{Stardust:sec:Motivation}

In this section, we first present a motivating disaster response scenario for the use of the 3D Continuum and, subsequently, we define requirements for an extensible next-generation simulator for the 3D Continuum.

%%%%%%%%%%%%%%%%%%%%%%%%%%%%%%%%%%%%%%%%%%%%%%%%%%%%%%%%%%%%%%%%%%%%%%%%%%%%%%%%%%
\subsection{Motivating Scenario}

While there are various use cases for the 3D Continuum, some of the most compelling \hlRevB{ones involve running distributed AI using a combination} of observation data from \EO{} satellites \hlRevB{and in-situ data from terrestrial sensors in a compound AI scenario~\cite{CompoundAI_6G_3DContinuum2025}}.
\EO{} data is large in size, e.g, each of the ESA Sentinel~2 satellites produces about 1.5~TB of data per day~\cite{ESA_Sentinel2Ops, Sentinel2CLaunched2024}, while downlink speeds to ground stations are typically approximately 300~Mbps\cite{EDRS_Overview}.
Thus, especially in cases where \EO{} data must be processed quickly \hlRevB{and possibly be augmented with data from in-situ sensors}, it is beneficial to preprocess it in a cluster of \LEO{} satellites to gain insights more quickly.

\begin{figure}
    \centering
    \includegraphics[width=.99\linewidth]{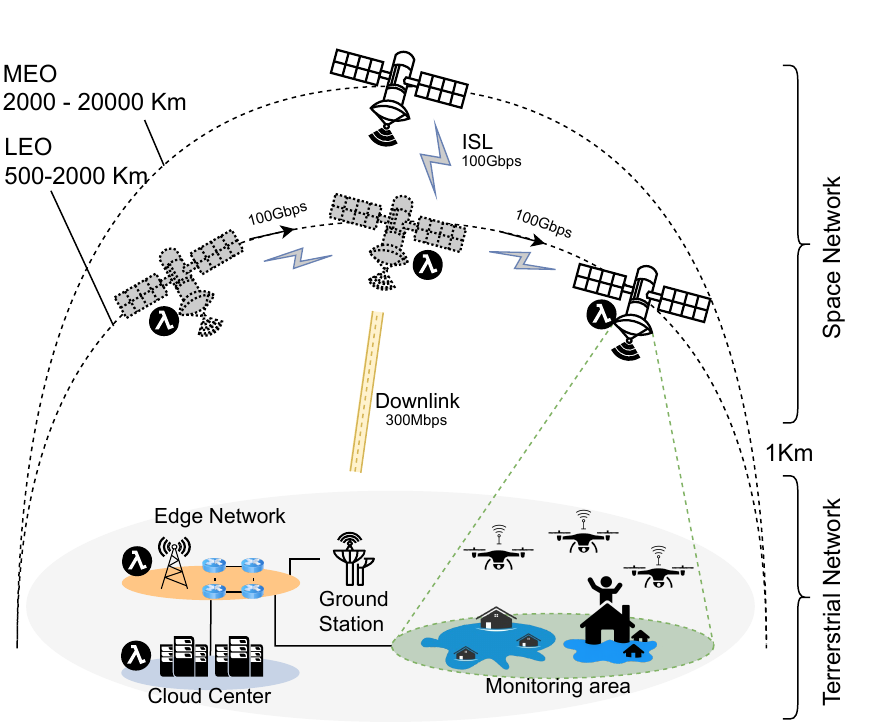}
    \caption{Motivating Use Case: Flood Disaster Response with Edge- and LEO-based Serverless Computing.}
    \label{Stardust:fig:MotivatingCase}
\end{figure}

Responding to natural disasters requires quick response times.
For example, if a suburban area is flooded after a hurricane, it is important to quickly identify people or animals that are in need of rescue.
\cref{Stardust:fig:MotivatingCase} shows a use case, where a combination of \EO{} satellites and drones is used to run a serverless workflow to find people and animals in need of help after a hurricane.
After the storm, drones fly over the affected area and record video data.
The drones are not powerful enough to run the ML model needed to detect people, so they need to offload this computation.
However, the cellular network has been damaged, so the video feed must be uplinked to \LEO{} satellites.
These \LEO{} satellites also receive data from an \EO{} satellite and combine that with the video from the drones to detect probable locations of survivors.
The identified locations are downlinked to a Cloud for detailed analysis and, \hlRev{if the presence of survivors is confirmed, are forwarded to rescue teams.}

Each of the serverless functions in this workflow has resource and latency requirements and must be placed on an appropriate node for execution.
This is done by a scheduler for the 3D Continuum.
Such schedulers are a hot research topic at the moment, and their evaluation requires a simulator that covers all node types of the 3D Continuum and provides node positions and network latencies.
Stardust allows such schedulers to be evaluated efficiently because its extensibility enables quick integration of the scheduling algorithms into the simulator.

\begin{figure}
    \centering
    \includegraphics[width=0.5\linewidth]{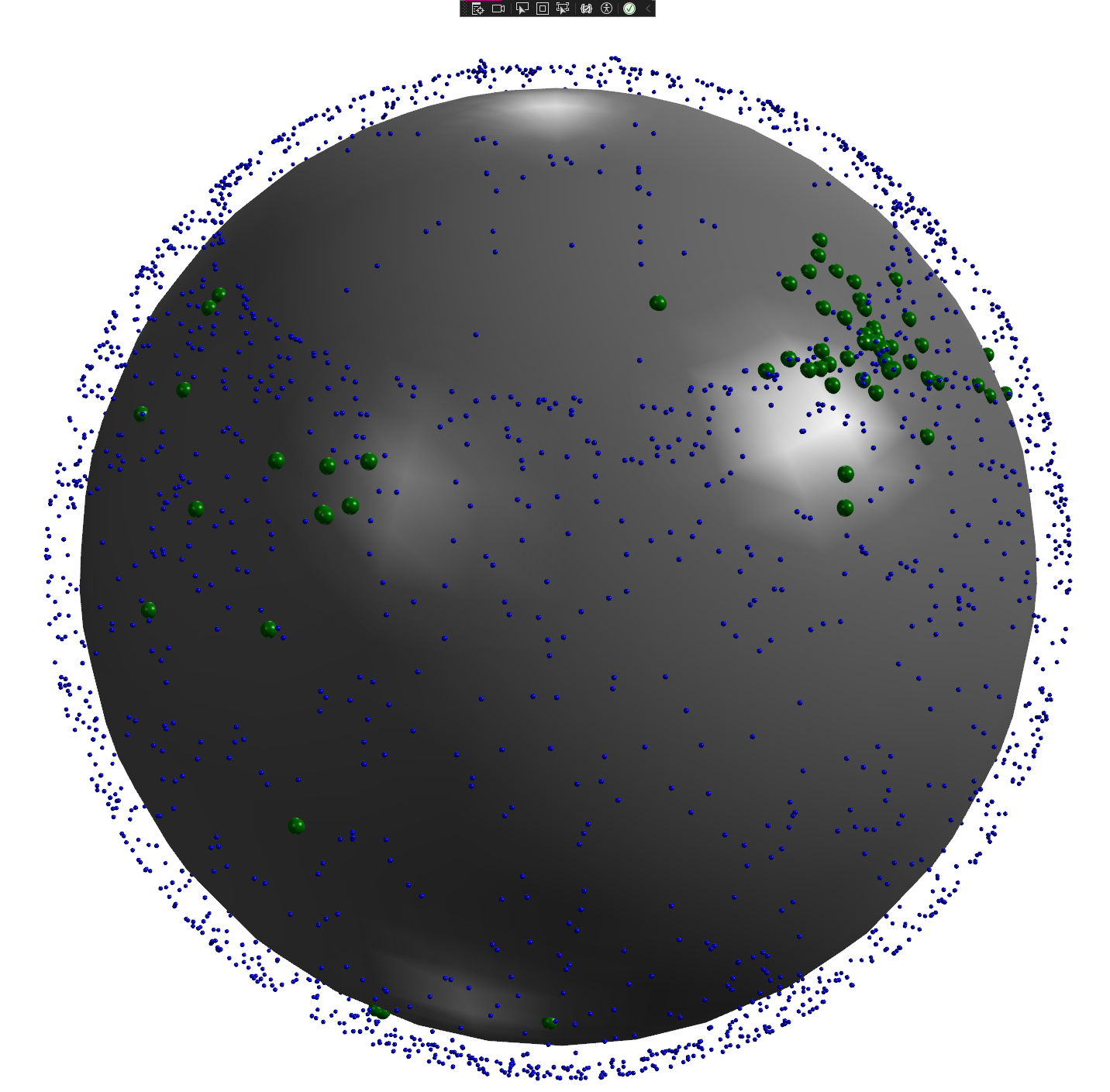}
    \caption{Visualization of the Simulated 3D~Continuum with Earth (gray), Satellites (blue), and Ground Stations (green). Ground Stations are Located in Selected Major Cities of Europe (right) and America (left).}
    \label{Stardust:fig:StardustVisualization}
\end{figure}

But Stardust is not limited to the evaluation of scheduling algorithms.
It can also be used for evaluating resource management, network routing, or other orchestration algorithms.
Additionally, Stardust will support the execution of workloads in the future, so the entire use case will be executable on the simulator.
\cref{Stardust:fig:StardustVisualization} shows a visualization of the 3D~Continuum simulated by Stardust with ground stations indicated in green and satellites shown in blue.

%%%%%%%%%%%%%%%%%%%%%%%%%%%%%%%%%%%%%%%%%%%%%%%%%%%%%%%%%%%%%%%%%%%%%%%%%%%%%%%%%%
\subsection{Requirements}

Based on the advantages and disadvantages of existing simulators/emulators, we define the following requirements for a next-generation simulator for the 3D Continuum:

\begin{enumerate}[label=R\arabic*]
    \item \emph{Simulate entire 3D Continuum}: The simulator must support simulating Edge, Cloud, and \LEO{} satellite nodes to allow evaluation of algorithms for the entire continuum.
    \label{Stardust:req:EntireContinuum}

    \item \emph{Configurable simulation steps}: The amount of simulated time that elapses in a single simulation step must be configurable, because some scenarios may require very fine-grained simulation steps (e.g., real-time use cases), while other scenarios need coarse-grained simulation steps for simulations that \hlRev{span multiple hours}.
    \label{Stardust:req:ConfigurableSteps}

    \item \emph{Extensibility}: The simulator must be easily extendable with custom logic to be executed at every simulation step.
    This enables the fast implementation of experiments for new algorithms.
    \label{Stardust:req:Extensibility}

    \item \emph{Information availability}: Custom logic must have access to all relevant data, such as node positions, node resources, and network routes.
    \label{Stardust:req:InfoAvailability}

    \item \emph{Choice between simulation and emulation}: Users must have the choice whether the experiments should execute as a simulation that tracks node positions, resources, and network state, and that executes custom logic or if selected nodes should be emulated to allow execution of workloads in containers or VMs.
    In emulation mode, only nodes that host a workload must execute a container or VM to keep resource usage to a minimum.
    \label{Stardust:req:SimulationEmulation}
    
\end{enumerate}

As we will explain in \cref{Stardust:sec:Design}, Stardust focuses on and fulfills \ref{Stardust:req:EntireContinuum}-\ref{Stardust:req:InfoAvailability}.
The complexity of \ref{Stardust:req:SimulationEmulation} merits a distinct in-depth evaluation.
Hence, we defer it to future work.

\section{Related Work}
\label{Stardust:sec:RelWork}

In this section, we discuss existing serverless platforms that explore workloads in the Edge-Cloud-Space 3D Continuum and \LEO{} simulators that enable the emulation and simulation of workflows in LEO edge constellations.

\subsection{LEO Edge Simulators}

Existing simulators for the 3D~Continuum can be divided into two main categories: network-only simulators and emulators.

\textbf{Network-only simulators} focus on simulating the network of a satellite mega constellation and their connections to ground stations, but they typically fail to account for the computational capabilities of \LEO{} satellites as nodes that can execute workloads.
The authors of~\cite{OpticalSatelliteSim2018} simulate satellite mega constellations with \ISLs{} and integrate them with ground stations, however, focusing purely on the network and not on workloads.
Hypatia~\cite{Hypatia2020} is a framework for simulating and visualizing the network behavior of LEO satellite constellations. It incorporates satellite-specific characteristics such as high-velocity orbital motion, \ISL{}, and ground-satellite links, enabling the evaluation of transport protocols, such as TCP and UDP, in a LEO-specific environment. 
Xeoverse~\cite{xeoverse2024} is a scalable and high-fidelity real-time simulation platform designed specifically for \LEO{} satellite mega-constellations. It models user terminals, satellites, and ground stations as lightweight VMs, pre-computing topology changes and focusing on relevant \ISL{} updates while streamlining link adjustments as needed. Xeoverse provides detailed network characteristics, including latency, capacity, signal-to-noise ratio (SNR), weather conditions, and antenna configurations.
\hlRev{
The popular and extensible network simulators ns-3~\cite{ns3Sim2010} and OMNeT++~\cite{OmnetPP2010} have also been used as bases for satellite simulators.
For example, SNS3~\cite{SNS3_2014} and ns-3-leo~\cite{ns3LEO_Swarm2022} are built on top of ns-3, while OS\textsuperscript{3}~\cite{OS3Sim2013} and its successor by Valentine and Parisis~\cite{OmnetPP_LEO2021} are based on OMNeT++.
All four enable the simulation of satellite and ground station networks with control over details such as network protocols, packets, and radio frequencies.
}

\textbf{Emulators} simulate the network of a satellite constellation and provision containers or VMs for the nodes of the 3D~Continuum to allow executing workloads on them.
However, emulators often suffer from limited scalability because the containers/VMs consume too many resources as the satellite constellation grows.
Celestial~\cite{Celestial2022} is a virtual testbed that emulates \LEO{} Edge satellite networks using microVMs. It precomputes satellite trajectories, bandwidth, and latencies between nodes at different points in time, allowing the orchestrator to manage network configuration requirements, such as SLOs, and to dynamically control microVMs based on the positions of the satellites.
However, Celestial lacks real-time satellite orbit positioning, as it relies on pre-calculated latencies.
Even though Celestial allows suspending microVMs, whose nodes are currently outside of a bounding box, e.g., the space above Europe, its approach is still resource-intensive for large satellite constellations, limiting the number of nodes that can be simulated on a single machine.
StarryNet~\cite{StarryNet2023} integrates real constellation data, including satellite trajectories, ground station distributions, and \ISL{} configurations, reproducing the spatial and temporal dynamics of mega-constellations while allowing researchers to deploy unmodified system code and simulate interactive network traffic. StarryNet ensures that its experiments reflect the scale and behavior of real-world \LEO{} networks, including time-varying connectivity and delays. StarryNet simulates constellations comprising thousands of satellites using a distributed, containerized setup across multiple machines. 

\hlRev{
Although these simulators and emulators offer space-ground integration, they have various shortcomings.
Simulators typically} focus on LEO-specific network simulation and do not account for the ability to execute workloads.
\hlRev{They also often implement many low-level networking details, which slow down large-scale simulations.
For example, the authors of the OS\textsuperscript{3} derivative~\cite{OmnetPP_LEO2021} report that a 5-minute simulation of 1,400~satellites with a step granularity of one second took more than four hours.
Emulators} focus on actual workload execution (not simulation), which makes them very resource-intensive, thus limiting their scalability for large satellite constellations\hlRev{, as we show in our experiments in \cref{Stardust:sec:EvalResults}.
}

\subsection{Evaluation Methods for LEO Platforms}
HyperDrive~\cite{HyperDrive2024} proposes a serverless platform that integrates devices across the Edge, Cloud, and space, creating a seamless continuum. HyperDrive enables serverless workflows to be executed across any layer within the 3D~Continuum. The scheduling mechanisms in HyperDrive consider processing capacities, such as CPU and memory, as well as specific properties of each layer, including satellite temperature and the battery levels of edge devices during the function placement process. However, HyperDrive relies on StarryNet's~\cite{StarryNet2023} network simulation to determine node positions, which does not include Edge devices.

Komet~\cite{Komet2024} introduces a serverless platform tailored for \LEO{} Edge computing, seamlessly integrating serverless functions with data replication to enable dynamic serverless function execution against satellite trajectories. By decoupling compute and state, Komet ensures virtual stationarity, allowing functions to maintain proximity to data despite the orbital movement of satellites. However, Komet's reliance on Celestial~\cite{Celestial2022} for network emulation focuses solely on satellite and ground station interactions, omitting broader integration with terrestrial Edge and Cloud nodes. 

% Krios~\cite{Krios2024}...

\begin{comment}
The idea is that other simulators are nice, but they have certain disadvantages:
\begin{itemize}
    \item They simulate nodes using containers or VMs, which makes them more heavyweight, meaning they cannot simulate as many nodes on a single machine.
    \item Some of them (e.g., Celestial) do not offer node position information during the iterations, because they just precalculate the latencies beforehand.
    \item Most support only \LEO{} and Cloud/ground station nodes, but no edge nodes.
    \item They are not easily extensible with custom logic.
\end{itemize}
\end{comment}

\section{Stardust Simulator Design}
\label{Stardust:sec:Design}

We now explain the architecture and core mechanisms of Stardust, which enable scalable simulations of the 3D~Continuum.

%%%%%%%%%%%%%%%%%%%%%%%%%%%%%%%%%%%%%%%%%%%%%%%%%%%%%%%%%%%%%%%%%%%%%%%%%%%%%%%%%%
\subsection{Stardust Architecture}

\begin{figure}
    \centering
    \includegraphics[width=.99\linewidth]{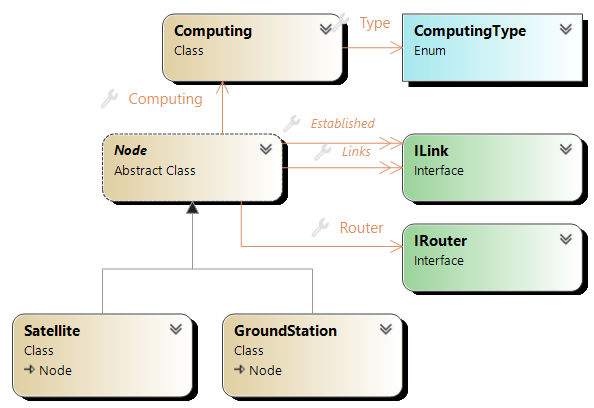}
    \caption{Stardust Simulator Core Abstractions.}
    \label{Stardust:fig:CoreAbstractions}
\end{figure}

Stardust is designed to use a modular architecture to enable extensibility.
% Using mainly interfaces and abstract classes, the structure of the simulator is defined. This abstraction makes it easy to adapt, extend, or replace individual components of the simulation.
The core abstractions are shown in \cref{Stardust:fig:CoreAbstractions}.
The central type from which all 3D Continuum nodes derive is \Node{}.

Every \Node{} has computational capabilities, communication links to other nodes, and the ability to route traffic. The abstract \Node{} class is currently implemented by the \inlinecode{GroundStation} and \inlinecode{Satellite} subclasses, required for different behaviors in movement. For \inlinecode{GroundStation}, a simplified movement (without inclined Earth axis and exact rotation duration) along the latitude every 24 hours is implemented. \inlinecode{Satellite} calculates the position by solving Kepler's equation for eccentric anomaly, considering only the gravitational interaction between the Earth and the satellite, \hlRev{i.e., it computes an unperturbed orbit.}
Since we focus on simulations of \hlRev{a few hours, this simplification helps} reduce the computational complexity and hardware requirements, while providing \hlRev{sufficient} accuracy.
\hlRev{Incorporation of perturbations, such as J2 and atmospheric drag~\cite{PerturbedKeplerOrbitFormulae2021}, is planned as future work.}
The \Computing{} class is responsible for tracking the available and used compute resources (e.g., CPU and memory) of each node. Each \Computing{} can be configured individually, such that simulated hardware can be set up to reflect specific conditions, e.g., no general-purpose compute hardware on older satellites, but GPU resources on the newest satellites. This can simulate the fact that the hardware of satellites in orbit cannot be repaired, replaced, or upgraded as easily as data center hardware on Earth.
\hlRev{To ensure realistic simulation of resource availability, the allocation of resources to tasks is handled as in Kubernetes, i.e., a task is assigned exclusive ownership of the requested resources for its entire execution duration.}
Each computing has a \inlinecode{ComputingType} which tags a Computing and further a Node and can potentially be used in routing, scheduling, or other algorithms. To create an Edge Node, an instance of \Node{}, i.e., either \inlinecode{Satellite} or \inlinecode{GroundStation}, gets assigned an instance of \inlinecode{Computing} tagged with \inlinecode{ComputingType.Edge}.

To simulate the available network connections \hlRev{with 100\% accuracy with respect to the node positions} in the 3D~Continuum, Stardust constructs a network graph that captures all nodes and the physical connections between them.
A direct physical connection between two nodes, e.g., a cable, radio, or laser link, is modeled by a link (\ILink{}) between those two nodes in the network graph.
Depending on the type of connection and other factors, such as distance, each link has particular latency and bandwidth properties.
Network routing relies on these links and their properties.
\IRouter{} is the abstraction for our dynamic routing mechanism, which may use either pre-route calculations, e.g., for algorithms that construct a routing table each step, or on-route calculations, for direct node-to-node routing like A\textsuperscript{*}.
% \ILink{} is an abstraction of links between nodes as links behave different. The currently implemented links might not have needed that abstraction, as both are communicating via air and links are constantly changing, but there might get other types of links implemented in future like simple cable wire connections on Earth or more complex links to moving plains or other satellite types (MEO, GEO, HEO). Routing relies on these links and their properties.

To allow exploring the behavior of the 3D~Continuum \hlRev{over multiple hours} within a reasonable timeframe, the speed at which time progresses in the simulation is configurable.
Within the simulation, time passes in discrete steps, called \emph{simulation steps}.
A simulation step can be configured to cover an arbitrary amount of time in the simulation, depending on the required granularity, e.g., one second, one minute, or five minutes.

\begin{algorithm}
\caption{Simulation Step Progression.}
\label{Stardust:alg:SimulationStep}
% \begin{small}
\begin{algorithmic}[1]
    \State \textbf{Input: } $t$: datetime; $N$: nodes; $P$: plugins
    
    \For{$n \in N$}
        \State \texttt{n.CalculatePosition(t)}
    \EndFor
    
    \For{$n \in N$}
        \State \texttt{n.UpdateLinks()}
    \EndFor

    \Statex \Comment{Optional/only for protocols with routing tables}
    \For{$n \in N$}
        \State \texttt{n.CalculateRoutingTable()}
    \EndFor

    \Statex \Comment{Run the plugins on step end}
    \For{$p \in P$}
        \State \texttt{p.PostSimulationStep()}
    \EndFor
\end{algorithmic}
% \end{small}
\end{algorithm}

At every simulation step, the state of the simulation is refreshed, i.e., the node positions and the network graph are updated.
% A simulation step represents a single iteration, which updates the simulation state, so all its nodes and their components will be set or calculated to match a realistic environment, depending on a given date time called simulation time. The simulation time can be specified at each step by an absolute date time or as a relative value in seconds after the previous simulation time.
% The concept behind a simulation step is described in \cref{Stardust:alg:SimulationStep}, which details the abstract functionalities required by the components. In the simulation, the for loops are parallelized.
\cref{Stardust:alg:SimulationStep} shows the high-level progression of a simulation step.
\begin{enumerate}
    \item Positions are calculated within the subclasses of \Node{}. Current implementations of \inlinecode{Satellite} and \inlinecode{Ground\\-Station} handle their distinct movement: orbiting Earth and points on Earth rotating around its axis. The resulting positions are Earth-centric coordinates to get unified positions.

    \item After all unified node positions are calculated, the physical links between nodes are established to update the network graph.
    % a link protocol can establish links between nodes, resulting in a network graph.
    % Ideally the graph is a single connected component, but, e.g., for small node numbers, there are might nodes which have no reachable neighbors to connect. This can lead to more than one connected component and not a single global connecting network.
    \label{Stardust:SimSteps:UpdateLinks}

    \item Using established links, a routing protocol can either calculate routing tables (e.g., using Dijkstra's algorithm) for constant route lookup times or, alternatively, routes can be calculated on-demand (e.g., a protocol using the A\textsuperscript{*} algorithm).
    \label{Stardust:SimSteps:CalcRoutingTable}
\end{enumerate}

Steps~\ref{Stardust:SimSteps:UpdateLinks} and~\ref{Stardust:SimSteps:CalcRoutingTable} are central to the network simulation, as we will discuss in the next subsection.

%%%%%%%%%%%%%%%%%%%%%%%%%%%%%%%%%%%%%%%%%%%%%%%%%%%%%%%%%%%%%%%%%%%%%%%%%%%%%%%%%%
\subsection{Dynamic Link Protocols and Routing Mechanism}

Simulating network communication in the 3D~Continuum consists of two steps, which can be realized in various ways.
To enable experimentation with different algorithms, Stardust relies on a dedicated abstraction to encapsulate the algorithm of each step:

\begin{enumerate}
    \item A \emph{link protocol} determines which pairs of nodes have a direct physical connection, such as a cable-, radio-, or laser link (i.e., physical and data-link layers of the OSI model).
    These are links in the network graph, each with a bandwidth and latency.

    \item A \emph{routing protocol} finds a route through the network graph for two nodes that want to communicate with each other (i.e., routing on the network layer of the OSI model).
    These are simple paths through the network graph, with bandwidths and latencies determined by the links along the path.
\end{enumerate}

Currently, Stardust supports the following link protocols: \inlinecode{mst} (Minimum Spanning Tree), \inlinecode{mst\_loop}, \inlinecode{mst\_smart\_loop}, \inlinecode{pst} (Parallel Spanning Tree), \inlinecode{pst\_loop}, and \inlinecode{pst\_smart\_loop}. A spanning tree ensures that the network graph forms a single connected component. The MST protocol runs Kruskal's algorithm on a single core to find the minimum spanning tree in the current constellation. The \inlinecode{pst} protocol filters for links that are eligible to connect and sorts links of satellites in parallel for pre-processing just before building the spanning tree. The \inlinecode{\_loop} suffix indicates that nodes with few links also add loops to the closest other nodes with few links. The \inlinecode{smart\_loop} variant adds loops to nodes with few links as well, but it attempts to find links that are in the opposite direction (relative to links previously established by the spanning tree) of existing links.
The smart loop variant chooses these additional links to approximate a +Grid-like structure at those nodes.
For each step, the \inlinecode{Ground\\-Satellite\\-Nearest\\-Protocol} establishes a link from a ground station to its nearest satellite.
Inter-satellite and ground-satellite link protocols run in parallel.

Routing protocols operate based on the established links and provide the latency of the shortest path through the built network graph. Currently, there are two \IRouter{} implementations: \inlinecode{DijkstraRouter} supports pre-route calculations, so the step calculation includes the calculation of the routing tables per node. The routing table gets filled using Dijkstra's algorithm. When routing information is requested, only a simple routing table lookup is required to obtain the result.
\inlinecode{AStarRouter} does not support pre-routing calculations, as it is a point-to-point path search algorithm. On a requested route, an A\textsuperscript{*} algorithm searches for the shortest path to the target node or service.

%%%%%%%%%%%%%%%%%%%%%%%%%%%%%%%%%%%%%%%%%%%%%%%%%%%%%%%%%%%%%%%%%%%%%%%%%%%%%%%%%%
\subsection{SimPlugin Extensibility}

To enable experimentation with different orchestration algorithms for the 3D~Continuum, Stardust allows plugging custom code into simulations as \emph{SimPlugins}.
A SimPlugin is a lightweight mechanism to execute custom code at the beginning and at the end of every simulation step.
To this end, a SimPlugin has full access to the simulation state, such as node positions, resources, and the network graph.
Additionally, it can leverage simulator services to compute network routes on demand and deploy (simulated) workloads on nodes.
% Before and after a simulation step, the simulation state can be inspected (view properties of nodes (position, computing, established links), network, computing,...), simulate some routes or used to schedule workloads.

\begin{figure}
    \centering
    \includegraphics[width=0.99\linewidth]{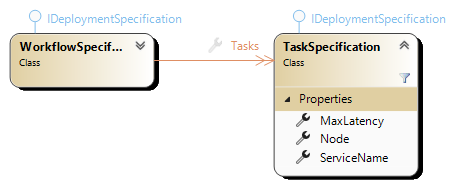}
    \caption{Class Diagram of \inlinecode{Workflow\\-Specification} with a  \inlinecode{Task\\-Specification}. \inlinecode{Task\\-Specification} has a max Latency to either a Node or a Service Name Configured as a Requirement.}
    \label{Stardust:fig:WorkflowClassDiagram}
\end{figure}

To deploy workloads in the simulation, there is the \inlinecode{IDeployment\\-Orchestrator} interface. There can be any number of such implementations registered to offer maximum flexibility for deployment algorithms and workloads. The implementation can access the simulation or other components by dependency injection in the constructor. A resolver delegates the requested workloads to the appropriate implementation that matches the workload requirements. \cref{Stardust:fig:WorkflowClassDiagram} shows that a \inlinecode{Workflow\\-Specification} consists of a list of \inlinecode{Task\\-Specifications}.
%% Thomas: I don't know if what I wrote in the next few sentences is correct, because I don't know the implementation well enough.
\hlRev{Each \inlinecode{Task\\-Specification} can be configured with a max latency SLO, which, by default, applies to the connection from the predecessor task.
However, the SLO can also be configured to refer to a specific service or node.}
Deploying a Workflow first resolves to the \inlinecode{Workflow\\-Orchestrator}, \hlRev{which is responsible for handling a \inlinecode{Workflow\\-Specification} and scheduling each of its tasks.
For each \inlinecode{Task\\-Specification} it finds the most suitable node and deploys the task there using the \inlinecode{Task\\-Orchestrator}.}

To create an additional orchestrator with new properties and scheduling strategy, e.g., to schedule a workload directly at the uplink satellite of a ground station, only a new class \inlinecode{Direct\\-Uplink\\-Orchestrator} implementing the \inlinecode{IDeployment\\-Orchestrator} interface, which handles instances of \inlinecode{Direct\\-Uplink\\-Specification} with a \inlinecode{Ground\\-Station} property, is needed. The new orchestrator has to be registered as a singleton to the \inlinecode{Host\\-Application\\-Builder} of the hosting framework. 

\cref{Stardust:lst:SimPlugin} shows an outline of a simple workload scheduler SimPlugin implementation.
All plugins must implement the \inlinecode{Sim\\-Plugin} interface, which provides handler methods for executing actions at the beginning and at the end of every simulation step.
\hlRev{As a scheduler, the plugin must also implement the \inlinecode{IDeployment\\-Orchestrator interface}}.
Using dependency injection, the program has access to \hlRev{simulator services, such as the simulation state and simulation controllers.
The simulation state allows inspecting all aspects of the simulation, e.g., the current node positions and the network graph.
The simulation controllers provide interfaces to modify the simulation state, e.g., deploy a task on a node.}
The \inlinecode{Simple\\-Scheduler} performs most work in \inlinecode{Post\\-Simulation\\-Step()}, where it dequeues the next task to be scheduled, then finds the most suitable node that fulfills the task's requirements, and, finally, deploys the task using the task orchestrator service of the simulation.

% \cref{Stardust:alg:SimPlugin} shows an outline of a SimPlugin implementation.
% The \inlinecode{IHosted\\-Service} interface must be implemented by all SimPlugins, which must then be registered with the \inlinecode{Host\\-Application\\-Builder} of the Stardust framework.
% Using dependency injection the program has access to the simulation controller or other registered components like simulator services, in case the program wants access to, e.g., inspect or modify some components like the simulation state. 

% \begin{figure}
%     \centering
%     \includegraphics[width=.99\linewidth]{figures/xeoverse-process.png}
%     % I commented that out bc it was giving a compiling error
%     \caption{Stardust Simulation Process. We could add a similar figure as a replacement for \cref{Stardust:alg:SimulationStep}. This is an example from Xeoverse}~\cite{xeoverse2024}
%     \label{Stardust:fig:Xeoverse}
% \end{figure}

% \begin{algorithm}
% \caption{Custom Stardust SimPlugin.}
% \label{Stardust:alg:SimPlugin}
% % \begin{small}
% \begin{algorithmic}[1]
%     \State \textbf{Input: } $sim$: Simulation; $N$: Nodes; $steps$: Integer

%     \Statex \Comment{simTime in seconds}
%     \State $simTime = Now$

%     \For{$i \in \{0..steps\}$}
%         \Statex \Comment{Do something before step}
%         \State \texttt{sim.Step(simTime)}
%         \Statex \Comment{Do something after step (e.g., deploy workloads, inspect nodes $N$)}
%         \Statex \Comment{Update simtime (e.g., +1min)}
%         \State $simTime = simTime + 60$
%     \EndFor
% \end{algorithmic}
% % \end{small}
% \end{algorithm}

\begin{minipage}{0.95\columnwidth}
\begin{codeListing}{Simple Workload Scheduler Stardust SimPlugin.}{Stardust:lst:SimPlugin}
public class SimpleScheduler 
    : ISimPlugin, IDeploymentOrchestrator {
  public SimpleScheduler(Simulation sim) {
    // Store sim and do 
    // other initialization.
  }
  
  public void PreSimulationStep(
      int stepIndex, DateTime simTime) {
    // Do work before the next step at the
    // specified simTime executes.
  }

  public void PostSimulationStep(
      int stepIndex, DateTime simTime) {
    TaskSpecification task = this.DequeueNextTask();

    if (task) {
      // If there is a new task to be
      // scheduled on this iteration,
      // find a target node that satisfies
      // the task's requirements.
      // ...
    
      this.sim.TaskOrchestrator
          .deploy(task, targetNode);
    }
  }
}
\end{codeListing}
\end{minipage}

\section{Evaluation \& Implementation}
\label{Stardust:sec:Evaluation}

In this section, we evaluate Stardust by integrating a simple network SLO-aware scheduler for the 3D Continuum as a SimPlugin and then conducting scalability experiments with the simulator.
All code required to run the experiments is part of our open source repository\footnote{\StardustRepoUrl{}}.

%%%%%%%%%%%%%%%%%%%%%%%%%%%%%%%%%%%%%%%%%%%%%%%%%%%%%%%%%%%%%%%%%%%%%%%%%%%%%%%%%%
\subsection{Implementation}

Stardust is implemented in C\# using the cross-platform .NET 8 framework.
The widely used \inlinecode{Microsoft.Extensions.Hosting} library simplifies the configuration and modularization of the application and facilitates the realization of the SimPlugin extensibility mechanism.
% It supports dependency injection through a standardized hosting model to achieve such easy modularization or adding SimPlugins.
Since each component is provided by dependency injection, replacing one component with another implementation only requires changing the hosting provider configuration.
This simplifies the process of switching implementations and configurations of the simulation and all components.

%%%%%%%%%%%%%%%%%%%%%%%%%%%%%%%%%%%%%%%%%%%%%%%%%%%%%%%%%%%%%%%%%%%%%%%%%%%%%%%%%%
\subsection{Experiment Design}

To evaluate Stardust, we mainly focus on its scalability, assessing its simulation performance and the performance of a simple workload scheduler plugin.
To this end, we implement a simple scheduler SimPlugin that places serverless workloads while fulfilling the resource requirements for every workload.
% This means that it ensures that each new function is scheduled on a node that has a network path with a latency less or equal to that of the SLO of the function.

We use a serverless workflow that is based on our flood disaster response use case from \cref{Stardust:sec:Motivation}.
It consists of four functions that are meant to be executed in sequence, each with distinct resource requirements.
% Each function has resource and latency requirements to its predecessor function.

The experiments are grouped into three categories.
A single iteration of an experiment comprises 100~simulation steps, with each simulation step covering one minute of simulated time.
\hlRev{These 100~minutes of simulated time are the period required for all satellites to complete one orbit around Earth, rounded up to the next multiple of 10.}
The used link protocol in Stardust is \inlinecode{pst\_smart\_loop}, unless otherwise noted.
The three experiment categories are the following:

\begin{enumerate}
    \item \emph{Simulator performance with respect to the infrastructure size}: We increase the total number of satellites for every experiment iteration, starting with 250~satellites and going up to 20k~satellites, which is about three times the current size of Starlink.
    The simulation runs without any SimPlugins to focus fully on the simulator performance.
    We measure the end-to-end runtime of every iteration and the system resource usage.
    Additionally, we compare Stardust's end-to-end runtime with Celestial~\cite{Celestial2022} and StarryNet~\cite{StarryNet2023}, two state-of-the-art \LEO{} mega constellation emulators.

    \item \emph{Scheduling performance with respect to the infrastructure size}: This experiment is the same as the first one, except that we schedule one workflow instance with our scheduler SimPlugin in every simulation step.

    \item \emph{Scheduling performance with respect to the workload}: We use our scheduler SimPlugin to deploy an increasing number of workflow instances per simulator step on a \LEO{} mega constellation consisting of 6,882~satellites.
    We measure the scheduling time on each simulation step.
\end{enumerate}

To set up realistic satellite orbits, we use TLE data on the orbits of 6,882~Starlink\footnote{\url{https://www.starlink.com}} satellites obtained from CelesTrack\footnote{\url{https://celestrak.org/NORAD/elements/}} on December~17, 2024.
For iterations that require more satellites, we duplicate existing satellites and offset their epoch, such that each duplicate is on the same orbital plane and altitude as the original satellite, but at a different position.
Since our focus is the satellites and since they require the most computational effort, because their positions need to be updated, we fix the number of ground stations (Clouds) to 85, which are distributed roughly equally across the globe.

% Throughout each of the experiment iterations, we monitor the resource usage of Stardust to assess its scalability.
All experiments are run on an Ubuntu 24.04~LTS VM with 32~CPU cores and 48~GiB RAM.
The underlying server is running an Intel Xeon processor of the Skylake generation.

%%%%%%%%%%%%%%%%%%%%%%%%%%%%%%%%%%%%%%%%%%%%%%%%%%%%%%%%%%%%%%%%%%%%%%%%%%%%%%%%%%
\subsection{Experimental Results}
\label{Stardust:sec:EvalResults}

\subsubsection{Simulator performance with respect to the infrastructure size}

We execute the experiment with seven satellite constellation sizes.
The end-to-end execution time of an experiment iteration consisting of 100~simulation steps is indicative of the simulator's performance, because it shows how well the satellite position computations and network graph updates scale.
Using the end-to-end execution time is also necessary to be able to compare Stardust to Celestial and StarryNet, because the latter two precompute all satellite positions before the simulation and launch a Firecracker microVM or a Docker container, respectively, for every node.
Additionally, they both advance simulation time in real time.
Thus, we configure them to simulate 100~seconds instead of minutes to avoid prolonging the end-to-end time artificially.
Celestial allows saving computational resources by suspending the microVMs of all satellites that are currently outside of a bounding box.
We use Europe as the bounding box, like in the example configuration supplied by Celestial

\begin{table}
\caption{Simulators End-to-end Results}
\label{Stardust:tab:e2eExperimentResults}
\centering
\resizebox{0.8\columnwidth}{!}{%
\begin{tabular}{lrrrl@{}}
\toprule
\textbf{Sat Count} & \textbf{Stardust (sec)} & \textbf{StarryNet (sec)} & \textbf{Celestial (sec)} \\ \midrule
250  & 3  & 503.67  & 115 \\
\rowcolor[gray]{0.9} 500  & 3  & 1400.33  & 126 \\
1k & 15  & 3457.67  & 163 \\
\rowcolor[gray]{0.9} 2k & 69  & -  & 328 \\
3k & 162  & -  & 673 \\
\rowcolor[gray]{0.9}6.8k & 877  & -  & - \\
13.8k & 2995 & - & - \\
\rowcolor[gray]{0.9}20.6k & 6464  & -  & - \\
\bottomrule
\end{tabular}%
}
\end{table}
\begin{figure}
    \centering
    \begin{tikzpicture}[font=\footnotesize]
      \begin{axis}[
        xlabel={Total Sats},
        xticklabel style={rotate=45},
        ylabel={E2E Time (sec)},
        legend style={at={(0.5,1.35)}, anchor=north, draw=none, legend columns=-1,font=\footnotesize},
        ymajorgrids=true,
        xmajorgrids=true,
        grid style=dashed,
        xticklabels={250, , , 2k, 3k, 6.9k, 13.8k, 20.6k},
        xtick=data,
        height=4cm,
        width=8.5cm,
        xmin=0.25,
        xmax=21,
        name=mainplot
      ] 
    
       % Plot for E2E Time PST
            \addplot[color=blue, mark=triangle*,mark size=1.5] coordinates {
                (0.25, 3) %250
                (0.5, 3) % 500
                (1, 15) % 1k
                (2, 69) %2k
                (3, 162) %3k
                (6.9, 877) % 6.9k
                (13.8, 2995) % 13.8k
                (20.6, 6464) %20.6k  
    
            };
            \addlegendentry{Stardust}
    
            % Plot for E2E Time Starrynet
            \addplot[color=red, mark=*,mark size=1.5] coordinates {
                (0.25, 503.67) %250
                (0.5, 1400.33) % 500
                (1, 3457.67) % 1k
    
            };
            \addlegendentry{Starrynet}
    
            % Plot for E2E Time Celestial
            \addplot[color=green!60!black, mark=square*,mark size=1.5] coordinates {
                (0.25, 115) %250
                (0.5, 126) % 500
                (1, 163) % 1k
                (2, 328) % 2k
                (3, 673) % 3k
    
            };
            \addlegendentry{Celestial}
    
      \end{axis}
      % Inset plot using nodes
        \node[anchor=north, inner sep=0pt] at ([yshift=0.35cm,xshift=-1.2cm]mainplot.north) {%
                \begin{tikzpicture}
                    \begin{axis}[
                        width=4.5cm,
                        height=3cm,
                        xmax=3,
                        ymax=750,
                        xticklabels={,,1k,2k,3k},
                        ytick={100,400,650},
                        grid style=dashed,
                        axis background/.style={fill=white},
                        x tick label style={yshift=-0.2em},
                        y tick label style={xshift=0.2em},
                        ytick pos=right
                    ]

                \addplot[color=blue, mark=triangle*,mark size=1] coordinates {
                    (0.25, 3) %250
                    (0.5, 3) % 500
                    (1, 15) % 1k
                    (2, 69) %2k
                    (3, 162) %3k
                    (6.9, 877) % 6.9k
                    (13.8, 2995) % 13.8k
                    (20.6, 6464) %20.6k  
                };
        
                % Plot for E2E Time Starrynet
                \addplot[color=red, mark=*,mark size=1] coordinates {
                    (0.25, 503.67) %250
                    (0.5, 1400.33) % 500
                    (1, 3457.67) % 1k
        
                };
        
                % Plot for E2E Time Celestial
                \addplot[color=green!60!black, mark=square*,mark size=1] coordinates {
                    (0.25, 115) %250
                    (0.5, 126) % 500
                    (1, 163) % 1k
                    (2, 328) % 2k
                    (3, 673) % 3k
        
                };
        
                \end{axis}
                \end{tikzpicture}
            };

    \end{tikzpicture}
    \caption{Experiment End-to-End Runtimes for Total Satellite Counts.}
    \label{Stardust:fig:E2EPerf}
\end{figure}
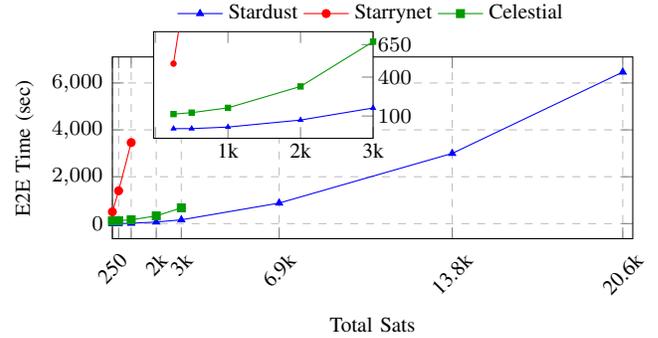

\cref{Stardust:fig:E2EPerf} compares the end-to-end experiment execution time of the simulators; the detailed results are in \cref{Stardust:tab:e2eExperimentResults}.
Due to the memory consumption of the microVMs, Celestial crashes during the iteration with 6.8k~satellites when run on a single machine.
StarryNet fails at the experiment with 2k~satellites on a single machine, because Docker is limited to attaching at most 1,024 virtual Ethernet adapters to a network bridge.
Thus, Celestial executes only the iterations up to 3k~satellites and StarryNet up to 1k~satellites.
Since Stardust does not execute any VMs or containers for the simulated nodes, it can handle much larger scenarios.
For better readability, \cref{Stardust:fig:E2EPerf} is split after 2k~satellites.
Even for small scenarios, Stardust executes experiments much faster than StarryNet.
For 250, 500, and 1,000~satellites, Stardust takes 3, 3, and 15~seconds, respectively.
Celestial requires 115, 126, and 163~seconds, respectively, whereas StarryNet needs 504, 1,400, and 3,457~seconds.
We observe that the Firecracker microVM setup and teardown of Celestial is much faster than the respective Docker actions of StarryNet, which may, however, be attributed to the fact that StarryNet sends each command through an SSH tunnel, while Celestial uses its own protobuf protocol.
Since Celestial and StarryNet progress the simulation in real time, 100~seconds is the lower bound for their end-to-end time.
But due to state precomputation and microVM/container setup/teardown, even the rest of StarryNet's execution time significantly exceeds that of Stardust and also grows faster than Stardust's.
For example, for 3k~satellites Celestial needs 673~seconds in total, while Stardust only needs 162~seconds.
The difference cannot only be attributed to microVM operations, because already the precomputation of the simulation state takes Celestial 432~seconds for this iteration.
This suggests that Stardust is more efficient at computing the simulation state, e.g., Celestial and StarryNet precompute the latencies between all pairs of nodes.
Stardust computes routes and latencies between a particular pair of nodes only when requested by a SimPlugin.
Altogether, Stardust's end-to-end experiment execution time scales \hlRev{with log-linear complexity} up to the largest experiment with 20,646~total nodes.

\begin{figure}
    \begin{tikzpicture}[font=\footnotesize]
        \begin{axis}[
            xlabel={Total Sats},
            ylabel={Exec Time (sec)},
            legend style={draw=none, font=\footnotesize},
            legend pos=north west,
            xticklabels={250, 500, 1k, 2k, 3k, 6.9k, 13.8k, 20.6k},
            xtick=data,
            grid=major,
            grid style=dashed,
            height=4cm,
            width=9cm,
            xmin=2,
            xmax=21,
            ymax=95
        ]
        % Plot for Mean Step Time MST
        \addplot[color=orange, mark=square*] coordinates {
            (0.25, 0.01351) %250
            (0.5, 0.04233) % 500
            (1, 0.19054) % 1k
            (2, 0.90098) %2k
            (3, 2.05638) %3k
            (6.9, 10.94533) % 6.9k
            (13.8, 39.1484) % 13.8k
            (20.6, 88.33054) %20.6k 
        };
        \addlegendentry{MST}

        % Plot for Mean Step Time PST
        \addplot[color=blue, mark=triangle*] coordinates {
            (0.25, 0.01586) %250
            (0.5, 0.03235) %500
            (1, 0.13001) %1k
            (2, 0.64393) %2k
            (3, 1.50536) %3k
            (6.9, 8.08111) %6.9k
            (13.8, 26.87099) %13.8k
            (20.6, 56.71087) %20.6k
        };
        \addlegendentry{PST}
        
        % Plot for Overall Average
        % \addplot[color=black, mark=x] coordinates {
        %     (0.25, 0.014685) %250
        %     (0.5, 0.03734) %500
        %     (1, 0.160275) %1k
        %     (2, 0.772455) %2k
        %     (3, 1.78087) %3k
        %     (6.9, 9.50322) %6.9k
        %     (13.8, 33.009695) %13.8k
        %     (20.6, 72.520705) %20.6k
        % };
        % \addlegendentry{Overall (avg)}
        \end{axis}
    \end{tikzpicture}
    \caption{Stardust Mean Execution Time for Single Simulation Step.}
    \label{Stardust:fig:MeanSimStepTime}
\end{figure}
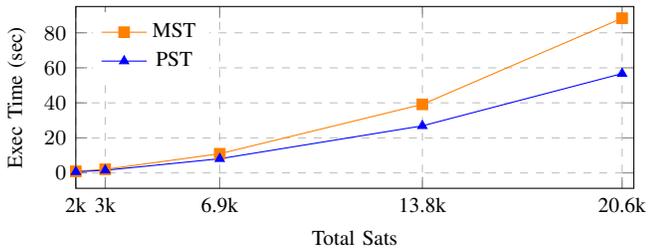

\cref{Stardust:fig:MeanSimStepTime} analyzes the mean execution time of a single simulation step in Stardust, executing with the \inlinecode{mst\_smart\_loop} and \inlinecode{pst\_smart\_loop} link protocols.
Both scale \hlRev{quadratically}, with PST having a more gentle slope.
For 20k~total satellites, a simulation step takes approximately 88~seconds with MST and about 57~seconds with PST.
The reason for PST not being even faster is that only the sorting part of Kruskal's algorithm is parallelized, indicating an avenue to improvement in the future.

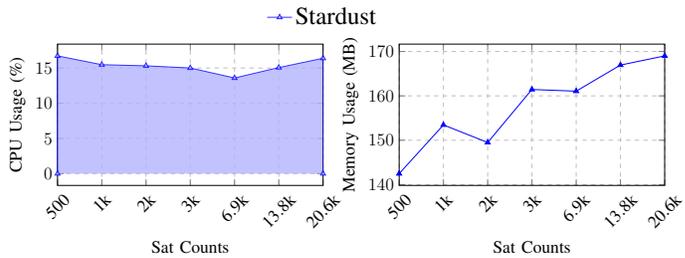
\begin{figure}[t]
\centering
\captionsetup[subfigure]{justification=centering}
\begin{subfigure}[t]{.23\linewidth}\centering
\begin{tikzpicture}[scale=.55,font=\large]
  \begin{axis}[
    xlabel={Sat Counts},
    xticklabel style={rotate=45},
    ylabel={CPU Usage (\%)},
    legend style={at={(1,1.3)}, anchor=north, draw=none, legend columns=-1,font=\LARGE},
    ymajorgrids=true,
    xmajorgrids=true,
    grid style=dashed,
    xticklabels={500, 1k, 2k, 3k, 6.9k, 13.8k, 20.6k},
    xtick=data,
    height=5cm,
    width=8cm,
    xmin=1,
    xmax=7,
  ] 

   \addplot[fill=blue!30, draw=blue, mark=triangle*, opacity=0.8]  coordinates
      {  
        % (1, 0) % to close the area
        % (1, 21.21) 
        % (2, 16.72) 
        % (3, 15.47) 
        % (4, 15.31) 
        % (5, 15.00) 
        % (6, 13.58) 
        % (7, 15.06) 
        % (8, 16.38) 
        % (8, 0) % to close the area

        %% Without x=250 to avoid the high CPU usage value we had there.
        (1, 0) % to close the area
        (1, 16.72) 
        (2, 15.47) 
        (3, 15.31) 
        (4, 15.00) 
        (5, 13.58) 
        (6, 15.06) 
        (7, 16.38) 
        (7, 0) % to close the area
    };
    \addlegendentry{Stardust}

  \end{axis}
\end{tikzpicture}
% \label{fig:}
\end{subfigure}
\hspace*{\fill}%
\begin{subfigure}[t]{.23\linewidth}\centering
\begin{tikzpicture}[scale=.55,font=\large]
  \begin{axis}[
    xlabel={Sat Counts},
    xticklabel style={rotate=45},
    ylabel={Memory Usage (MB)},
    legend pos=north west,
    legend style={draw=none},
    ymajorgrids=true,
    xmajorgrids=true,
    grid style=dashed,
    xticklabels={500, 1k, 2k, 3k, 6.9k, 13.8k, 20.6k},
    xtick=data,
    height=5cm,
    width=8cm,
    xmin=2,
    xmax=8,
  ] 
  \addplot[blue,mark=triangle*] coordinates
      {
        %% Without x=250 to avoid the high CPU usage value we had there.
        % (1, 133.72) 
        (2, 142.47) 
        (3, 153.45) 
        (4, 149.49) 
        (5, 161.44) 
        (6, 161.06) 
        (7, 166.94) 
        (8, 169.01) 
      };
  \end{axis}
\end{tikzpicture}
% \label{fig:}
\end{subfigure}
\hspace*{\fill}%
\caption{\hlRev{Stardust Resource Usage for Total Satellite Counts.}}
\label{Stardust:fig:ResUsage}
\end{figure}

\cref{Stardust:fig:ResUsage} shows the mean resource usage of Stardust \hlRev{for 500 to 20k satellites}.
The mean CPU usage never exceeds 25\% and memory usage goes up to 170~MB for 20k~satellites.
The resource usage for Celestial and StarryNet is not comparable, because they execute microVMs or containers for nodes, e.g., the 6.8k~satellites experiment failed with Celestial, because the host's 48~GB of RAM was exhausted.
Stardust intentionally does not execute nodes, hence, it can simulate much larger constellations on a single machine.

Stardust's resource usage in combination with the \hlRev{quasilinear} scalability of the end-to-end experiment runtime indicates that it can scale to even larger constellation sizes on a single machine.
Thus, it is well suited for the upcoming expansions of \LEO{} mega constellations.

\subsubsection{Scheduling performance with respect to the infrastructure size}

In this experiment, we execute the same iterations as for the previous one, but, additionally, we use our scheduler SimPlugin to deploy one instance of the flood disaster response workflow, i.e., four functions, on every simulation step.
We execute the experiment with both link protocols, i.e., \inlinecode{mst\_smart\_loop} and \inlinecode{pst\_smart\_loop}.
Since Celestial does not support plugins and StarryNet's node model does not account for compute resources, we run this experiment only with Stardust.

The goal of Stardust's SimPlugins is to provide a lightweight mechanism for integrating custom code into the simulation, such as an orchestration or scheduling algorithm that should be validated.
Celestial provides no plugin mechanism to extend the simulation, but it offers a REST API that is accessible from within each microVM.
This API allows querying network route information between two nodes and provides rudimentary information about the nodes.
However, there is no way to query a node's resources and location or to deploy a workload in the 3D~Continuum -- this would require installing a full-fledged orchestrator, such as Kubernetes, in the scenario.
StarryNet can be used as a library in a Python script to design and run a custom scenario.
However, its API is limited, e.g., its node model has no concept of computational resources, which makes it difficult to write orchestration algorithms.
Stardust's SimPlugins are a lightweight mechanism for executing custom algorithms directly as part of the simulation.
SimPlugins have full access to the entire simulation state, which includes the compute resources on nodes and locations, and provides APIs to query network routes and deploy simulated workloads on the nodes.
This enables SimPlugins to be used for evaluating orchestration algorithms without using a full-fledged orchestrator, like Kubernetes, which would introduce additional restrictions to the simulation.

%% Experiment 2: Schedule 1 workload on increasing infra size
\begin{figure}
    \begin{tikzpicture}[font=\footnotesize]
        \begin{axis}[
            xlabel={Total Sats},
            ylabel={Scheduling Time (ms)},
            legend style={draw=none, font=\footnotesize},
            legend pos=north west,
            xticklabels={250, 500, 1k, 2k, 3k, 6.9k, 13.8k, 20.6k},
            xtick=data,
            grid=major,
            grid style=dashed,
            height=4cm,
            width=9cm,
            xmin=2,
            xmax=21,
            % ymode=log
        ]
        % Plot for Mean Step Time MST
        \addplot[color=orange, mark=square*] coordinates {
            (0.25, 1.23) % 250
            (0.5, 3.32) % 500
            (1, 5.33) % 1k
            (2, 8.7) % 2k
            (3, 13.54) % 3k
            (6.9, 30.12) % 6.9k
            (13.8, 79.59) % 13.8k
            (20.6, 144.44) % 20.6k
        };
        \addlegendentry{MST}

        % Plot for Mean Step Time PST
        \addplot[color=blue, mark=triangle*] coordinates {
            (0.25, 1.53) % 250
            (0.5, 2.91) % 500
            (1, 4.85) % 1k
            (2, 7.76) % 2k
            (3, 13.09) % 3k
            (6.9, 32.39) % 6.9k
            (13.8, 73.82) % 13.8k
            (20.6, 137.31) % 20.6k
            
        };
        \addlegendentry{PST}
        
        \end{axis}
    \end{tikzpicture}
    \caption{Scheduling Performance for 1~Workflow on Increasing Constellation Sizes.}
    \label{Stardust:fig:SchedulingPerf}
\end{figure}
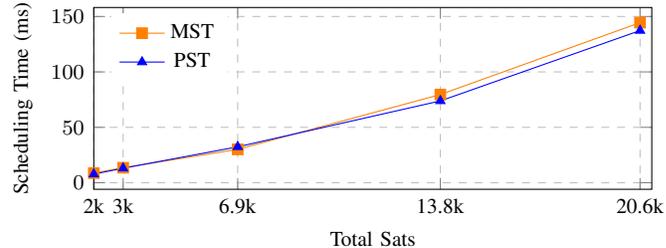

\cref{Stardust:fig:SchedulingPerf} illustrates the mean duration of scheduling one workflow instance for various satellite constellation sizes. 
This shows that the performance of the simulator state and deployment APIs scales \hlRev{almost linearly} with the infrastructure size and does not cause a bottleneck for SimPlugins.

\subsubsection{Scheduling performance with respect to the workload}

In the third experiment, we evaluate the speed and scalability of workload deployment operations in a Stardust SimPlugin.
The number of satellites is fixed to 6,882, i.e., the size of the Starlink constellation on December~17, 2024.
The simulation consists of 100~steps; in every step we schedule a fixed number of flood disaster response workflow instances.
We perform three iterations, with 1, 10, and 100~scheduled workflow instances (four functions per instance) per simulation step.
As for the previous experiment, due to the restrictions of Celestial and StarryNet, we run this experiment only with Stardust.

%% Lineplot for WorkloadScalabilitySchedTime with and without rescheduling
\begin{figure}
    \centering
    \begin{tikzpicture}[font=\footnotesize]
      \begin{axis}[
        xlabel={Workflow Instances},
        ylabel={Scheduling Time (ms)},
        legend style={draw=none, font=\footnotesize},
        legend pos=north west,
        ymajorgrids=true,
        xmajorgrids=true,
        grid style=dashed,
        xticklabels={1, 10, 100}, %, 250},
        xtick=data,
        height=5cm,
        width=8cm
        % ymode=log
      ] 
            % % Plot for Sched Time MST without rescheduling
            % \addplot[color=magenta, mark=square*] coordinates {
            %     (1, 43.17)
            %     (10, 70.62)
            %     (100, 558.39)
    
            %     % (1, 0.04317) %1
            %     % (2, 0.07062) % 10
            %     % (3, 0.55839) % 100
            % };
            % \addlegendentry{MST}
    
            % % Plot for Sched Time PST without rescheduling
            % \addplot[color=cyan, mark=triangle*] coordinates {
            %     (1, 385.4)
            %     (10, 1162.94)
            %     (100, 1098.85)
    
            %     % (1, 0.3854) %1
            %     % (2, 1.16294) %10
            %     % (3, 1.09885) %100
            % };
            % \addlegendentry{PST}
            
            % Plot for Sched Time MST with rescheduling
            \addplot[color=orange, mark=square*] coordinates {
                (1, 30.78)
                (10, 29.41)
                (100, 757.21)
                % (250, 2017.35)
                
                % (1, 0.03078) %1
                % (2, 0.02941) % 10
                % (3, 0.75721) % 100
                % (4, 2.01735) %250
            };
            \addlegendentry{MST} % w resched.}
    
            % Plot for SchedTime PST with rescheduling
            \addplot[color=blue, mark=triangle*] coordinates {
                (1, 33.17)
                (10, 66.89)
                (100, 736.19)
                % (250, 2166.57)
                
                % (1, 0.03317) %1
                % (2, 0.06689) %10
                % (3, 0.73619) %100
                % (4, 2.16657) %250
            };
            \addlegendentry{PST} % w resched.}
    
      \end{axis}
    \end{tikzpicture}
    \caption{Workload Scalability Experiment with 6,882~Satellites -- Scheduling Time.}
    \label{Stardust:fig:WorkloadScalabilitySchedTime}
\end{figure}
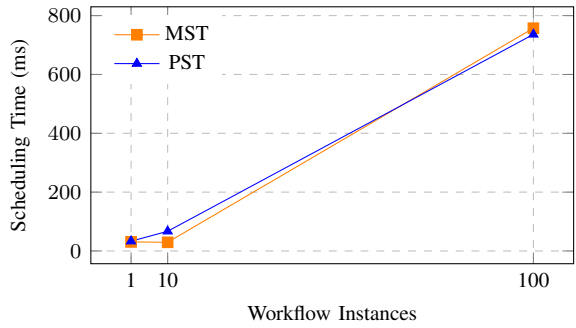

\cref{Stardust:fig:WorkloadScalabilitySchedTime} shows the mean execution times of the scheduler SimPlugin.
The time scales linearly from 33~ms when scheduling a single workflow instance per simulation step to 736~ms when scheduling 100~workflow instances per step.
This shows that the simulator state API does not present any bottleneck and that it is suitable for validating 3D~Continuum orchestration algorithms.

\begin{table}
\caption{Stardust Experiment Results}
\label{Stardust:tab:experimentResults}
\resizebox{\columnwidth}{!}{%
\begin{tabular}{p{1.5cm}lrrrrrl@{}}
\toprule
\textbf{Sat/Workload Count} & \textbf{Algo} & \textbf{Step (ms)} & \textbf{Sched (ms)} & \textbf{CPU (\%)} & \textbf{RAM (MB)} \\ \midrule
\multicolumn{6}{c}{\textbf{Simulator Performance}} \\ \midrule
\rowcolor[gray]{0.9} 250  & MST  & 13.51   &  -    & 7.19  & 143.41 \\
250  & PST  & 15.86   &  -    & 21.21 & 133.73 \\
\rowcolor[gray]{0.9} 3,023 & MST  & 2,056.38 &  -    & 8.92  & 152.38 \\
3,023 & PST  & 1,505.36 &  -    & 15.00 & 161.45 \\
\rowcolor[gray]{0.9} 20,646 & MST & 88,330.54 & -    & 9.05  & 166.40 \\
20,646 & PST & 56,710.87 & -    & 16.38 & 169.02 \\
\midrule
\multicolumn{6}{c}{\textbf{Scheduling Performance}} \\ \midrule
\rowcolor[gray]{0.9} 250  & MST  & 25.88  & 1.23  & 10.53  & 141.92 \\
250  & PST  & 24.66  & 1.53  & 10.76  & 137.49 \\
\rowcolor[gray]{0.9} 3,023 & MST  & 2,194.02 & 13.54 & 9.98   & 160.40 \\
3,023 & PST  & 1,605.45 & 13.09 & 14.96  & 160.15 \\
\rowcolor[gray]{0.9} 20,646 & MST & 87,905.50 & 144.44 & 9.35  & 184.88 \\
20,646 & PST & 57,598.34 & 137.31 & 16.47  & 165.52 \\
\midrule
\multicolumn{6}{c}{\textbf{Scheduling Workload Scalability}} \\ \midrule
\rowcolor[gray]{0.9} 1  & MST  & 10,598.31  & 30.78  & 9.76  & 167.36 \\
1  & PST  & 8,457.07  & 33.17  & 14.24  & 165.96 \\
\rowcolor[gray]{0.9} 10 & MST  & 10,716.2 & 29.41 & 9.58   & 166.25 \\
10 & PST  & 38,599.87 & 736.19 & 23.77  & 167.21 \\
\rowcolor[gray]{0.9} 100 & MST & 42,901.99 & 757.21 & 21.49  & 168.30 \\
100 & PST & 38,599.87 & 736.19 & 21.49  & 168.57 \\
\bottomrule
\end{tabular}%
}
\end{table}

\cref{Stardust:tab:experimentResults} provides a summarizing overview of all experiment results.

\section{Conclusion}
\label{Stardust:sec:Conclusion}

In this paper, we have presented Stardust, a scalable and extensible simulator for the 3D~Continuum.
Stardust aims to provide a platform for evaluating new orchestration algorithms for the 3D~Continuum on a single machine.
To this end, Stardust simulates all nodes of the 3D~Continuum, including the orbital movement of satellites, maintains a network graph, and allows progressing simulated time at a configurable pace for short-term or long-term simulations.
Stardust intentionally does not emulate the execution of workloads to allow simulating mega constellations up to three times the current size of Starlink, as shown in our experiments.

In the near future, we intend to \hlRev{implement approximation of perturbed orbits and} introduce emulator capabilities to Stardust to allow it to execute workloads using \hlRev{even more} realistic network routes, \hlRev{including QoS-based routing, in} the 3D Continuum.
Users will be able to select whether Stardust should operate in simulator mode or emulator mode.
We plan to introduce a lightweight emulator mode using containers and an orchestrated emulator mode, where Kubernetes can be used in the 3D~Continuum.
Our aim is to \hlRev{employ sparse execution of containers, i.e.,} only execute containers for nodes that have been assigned a workload, to maintain the scalability of Stardust.
A distributed operation mode across multiple machines, \hlRev{in combination with sparse container execution,} will continue to allow experiments with the growing \LEO{} mega constellations of the future.

% Balance the columns on the last page.
% Despite the warning, we call \balance after the inputs to avoid issues with footnotes floating inside the text.
\balance

\ifthenelse{\boolean{anonymousMode}}{}{
    \section*{Acknowledgment}
        This work is partially funded by the Austrian Research Promotion Agency (FFG) under the project RapidREC (Project No. 903884).
        This work has received funding from the Austrian Internet Stiftung under the NetIdee project LEO Trek (ID~7442).
        This research received funding from the EU's Horizon Europe Research and Innovation Program through the TEADAL (GA No. 101070186) and NexaSphere projects (GA No. 101192912).   
}
% References section
\bibliographystyle{IEEEtran}
\bibliography{IEEEabrv, bibliography}

\end{document}